\documentclass[fleqn,usenatbib]{mnras}

\usepackage{newtxtext,newtxmath}

\usepackage[T1]{fontenc}

\DeclareRobustCommand{\VAN}[3]{#2}
\let\VANthebibliography\thebibliography
\def\thebibliography{\DeclareRobustCommand{\VAN}[3]{##3}\VANthebibliography}

\usepackage{graphicx}	
\usepackage{amsmath}	
\usepackage{subfiles}
\usepackage{xcolor}
\usepackage{orcidlink}
\usepackage{float}

\usepackage{siunitx}
\DeclareSIUnit\year{yr}
\DeclareSIUnit\parsec{pc}
\DeclareSIUnit\msun{M_\odot}
\DeclareSIUnit\Rsun{R_\odot}

\newcommand{\kms}{\unit{\km\per\s}}
\newcommand{\kpc}{\unit{\kilo\parsec}}
\newcommand{\pc}{\unit{\parsec}}

\newcommand{\Lz}{L_{\rm z}}

\newcommand{\ignore}[1]{}

\usepackage{pifont}
\usepackage{soul}

\title[Angular momentum spiral of the Milky Way disc]{The angular momentum spiral of the Milky Way disc in Gaia}

\author[Yaaqib, Naik, Pe\~{n}arrubia, Petersen]{
Rashid Yaaqib,$^{1,2}$\thanks{E-mail: rashid.yaaqib@ed.ac.uk}\orcidlink{0009-0003-9063-1382}
Aneesh P. Naik,$^{1}$\orcidlink{0000-0001-6841-1496}
Jorge Pe\~{n}arrubia,$^{1}$
Michael S. Petersen$^{1}$\orcidlink{0000-0003-1517-3935}\\
$^{1}$Institute for Astronomy, University of Edinburgh, Royal Observatory, Blackford Hill, Edinburgh EH9 3HJ, UK\\
$^{2}$Department of Physics, United Arab Emirates University, Al Ain, Abu Dhabi, UAE\\
}
\date{Accepted XXX. Received YYY; in original form ZZZ}

\pubyear{2024}

\begin{document}
\label{firstpage}
\pagerange{\pageref{firstpage}--\pageref{lastpage}}
\maketitle

\begin{abstract}
Data from the {\it Gaia} mission shows prominent phase-space spirals that are the signatures of disequilibrium in the Milky Way (MW) disc. In this work, we present a novel perspective on the phase-space spiral in angular momentum (AM) space. Using {\it Gaia} DR3, we detect a prominent AM spiral in the solar neighbourhood. We demonstrate the relation of AM to the $z-v_z$ spiral and show that we can map to this space from angular momentum through simplifying assumptions. By modelling the orbit of stars in AM, we develop a generative model for the spiral where the disc is perturbed by a bulk tilt at an earlier time. Our model successfully describes the salient features of the AM spiral in the data. Modelling the phase spiral in AM is a promising method to constrain both perturbation and MW potential parameters. Our AM framework simplifies the interpretation of the spiral and offers a robust approach to modelling disequilibrium in the MW disc using all six dimensions of phase space simultaneously.
\end{abstract}

\begin{keywords}
Galaxy: general; Galaxy: kinematics and dynamics; Galaxy: disc; galaxies: interactions; Galaxy: solar neighbourhood
\end{keywords}


\section{Introduction}

Equilibrium models of the Milky Way (MW) have long been used as the interpretive base for increasingly large datasets of the MW stars. However, recent data from the \textit{Gaia} mission \citep{gaia_collaboration_gaia_2016, gaia_collaboration_DR3_2023} have revealed the dynamical disequilibrium of the MW disc in both the vertical \citep{antoja_dynamically_2018,huntMultiplePhaseSpirals2022,Alinder.spiral.2023,Antoja.spiral.2023} and radial \citep{antoja_tidally_2022,caoRadialWaveGalactic2024b,hunt_radial_2024} projections of six-dimensional phase space (e.g. in $z-v_z$, $R-v_r$ and $R-v_\phi$). The aforementioned projections are a non-exhaustive list of representations of disequilibrium patterns observed.

Perhaps the most striking feature of the data is the spiral pattern observed in the vertical coordinates $z-v_z$. In this phase-space projection, the data show a spiral pattern owing to incomplete phase mixing in the Galactic disc. The presence of the feature provides us with a unique opportunity to study disequilibrium processes in real-time and is proving to be a valuable source of information about the Galactic potential, perturbative events and the dynamical evolution of the MW \citep{widmarkWeighingGalacticDisk2022, liStellarDistributionFunction2021,widmarkMappingMilkyWay2022, darragh-fordESCARGOTMappingVertical2023,guoMeasuringMilkyWay2024}. 

The origin of the phase spiral has been linked to perturbations to the disc. However, there is no consensus on the nature of the perturbation. One hypothesis is that the passage of the Sagittarius dwarf galaxy through the disc plane seeds the perturbation, where the tidal forces generate vertical and radial oscillations in the disc \citep{binneyOriginGaiaPhaseplane2018,laporteFootprintsSagittariusDwarf2019}. Other proposed formation mechanisms include vertical oscillations triggered by bar buckling \citep{khoperskovEchoBarBuckling2019}, moving groups such as streams \citep{michtchenkoMovingGroupsOrigin2019} or other small-scale perturbations \citep{tremaineOriginFateGaia2023}, and long-lived perturbations in the MW dark matter halo \citep{Grand.snail.2023}.

One space often not considered when looking at the kinematic signatures of disequilibrium in the MW disc is angular momentum (AM). In axisymmetric equilibrium disc models, the AM component perpendicular to the plane, $L_z$, is a conserved quantity. Meanwhile, for perfectly planar orbits, $L_x, L_y=0$. However, the majority of disc stars exhibit some oscillation perpendicular to the plane; such stars trace oscillatory routes through $L_x-L_y$ space. When the epicycles of individual stars in the disc are not correlated, the resulting projection of $L_x-L_y$ is simply a `blob' centred on zero. There are several advantages to AM space. First, choosing only two of the three components uses the full six-dimensional phase space information of the stars, this is advantageous when choosing the coordinates in which the disequilibrium feature is modelled as it incorporates both radial and vertical terms, encompassing all the projections in which disc disequilibrium has been observed. Also, AM can be measured directly from observational data without integrating orbits in an assumed potential, in contrast to other integrals such as the actions. Finally, as we shall see in Section \ref{sec:theory}, the trajectories of individual stars in AM (particularly in cylindrical coordinates) take simple forms, allowing straightforward modelling of kinematic signatures such as the spiral.

In this paper we present a novel outlook on the phase space spiral observed in the \textit{Gaia} data through the AM lens. In Section \ref{sec:mainresult} we show the cylindrical AM spiral in the {\it Gaia} data in the local volume. Section \ref{sec:theory} lays the framework for interpreting the spiral, through consideration of the orbits of individual stars, then presents the spiral model used in the remainder of the work. The results of our modelling is discussed in Section \ref{sec:discussion}, where we address the insights gained through modelling in this space, and the limitations. Finally, we provide some concluding remarks in Section \ref{sec:conclusion}.
\begin{figure*}
    \begin{center}
\includegraphics{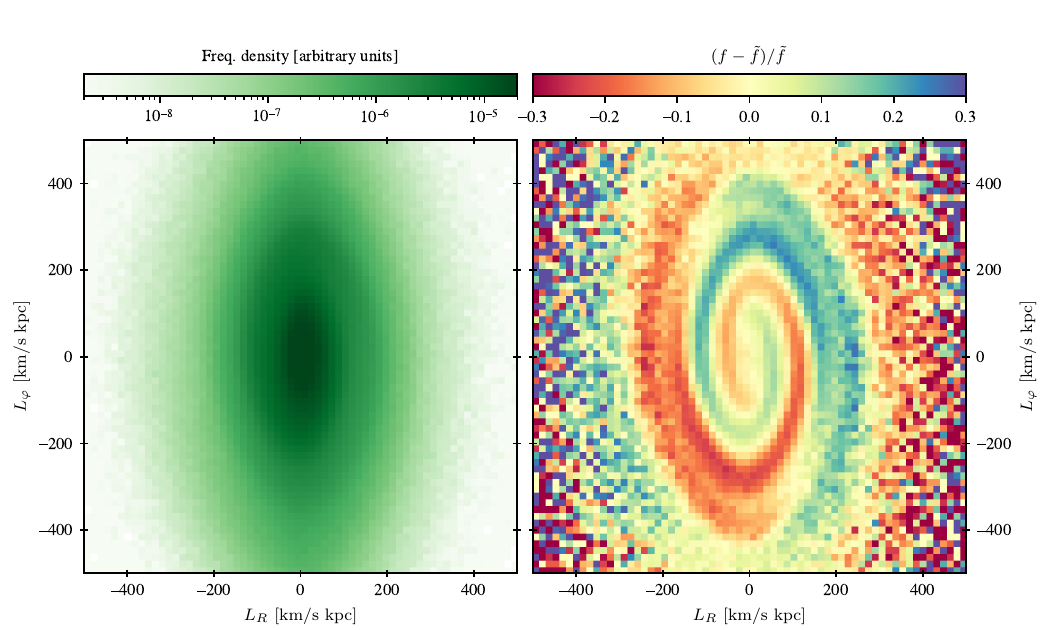}
    \caption{The \textit{Gaia} AM spiral in cylindrical coordinates. The left panel show the raw data histogram, while the right panels show residuals against a symmetric model. A spiral-shaped over-/under-density is clearly visible, at around the 20\% level.}
    \label{fig:observedspiral}
    \end{center}
\end{figure*}

\section{The Gaia RVS sample}
\label{sec:data}
To build the parent sample, we query the  {\it Gaia} DR3 \citep{gaia_collaboration_DR3_2023} catalogue, accepting stars that have radial velocity measurements, a reduced weighted unit error of {\tt ruwe}<1.4 and a relative parallax error of $\varpi/\sigma_{\varpi} >5$. Appendix \ref{appendix:query} gives our ADQL query. To transform the ICRS coordinates of the \textit{Gaia} stars to the Cartesian Galactocentric frame, we adopt a right-handed coordinate system with the Sun positioned at $\vec{r}_{\odot \rightarrow MW}=(-8.3, 0.0, 0.02)~\kpc$ \citep{gravitycollaborationImprovedGRAVITYAstrometric2021,bennett_vertical_2018}, with velocity $\vec{v}_{\odot \rightarrow MW}=(11.1, 244.24, 7.24)~\kms$ \citep{schonrich_local_2010,eilers_circular_2019}. Finally, the angular momenta of stars, used in subsequent sections, are computed using the positions and velocities of stars in the Galactocentric frame. The total number of stars in the parent sample is 26,747,826 stars, for the remainder of the work presented, we use only stars confined to the annular sector defined by the radial condition of $|R - 8.3~\kpc| < 1.0~\kpc$ and azimuthal condition of $|\varphi - \pi| < 1.0/8.3$ where $(R,\varphi)$ are the cylindrical galactocentric coordinates. The total number of stars after applying the cut is 11,827,676 stars, this is the main sample of stars used in all subsequent figures. Stars within this region have a mean AM uncertainty in each Cartesian direction of $\approx 30~\kms\kpc$.\footnote{Computed via Monte Carlo error propagation of the uncertainties on distance, proper motion and radial velocity as reported from the {\it Gaia} table. Note that proper motion correlations were accounted for during the propagation.}

\section{The angular momentum spiral}
\label{sec:mainresult}

\begin{figure}
    \begin{center}
    \includegraphics{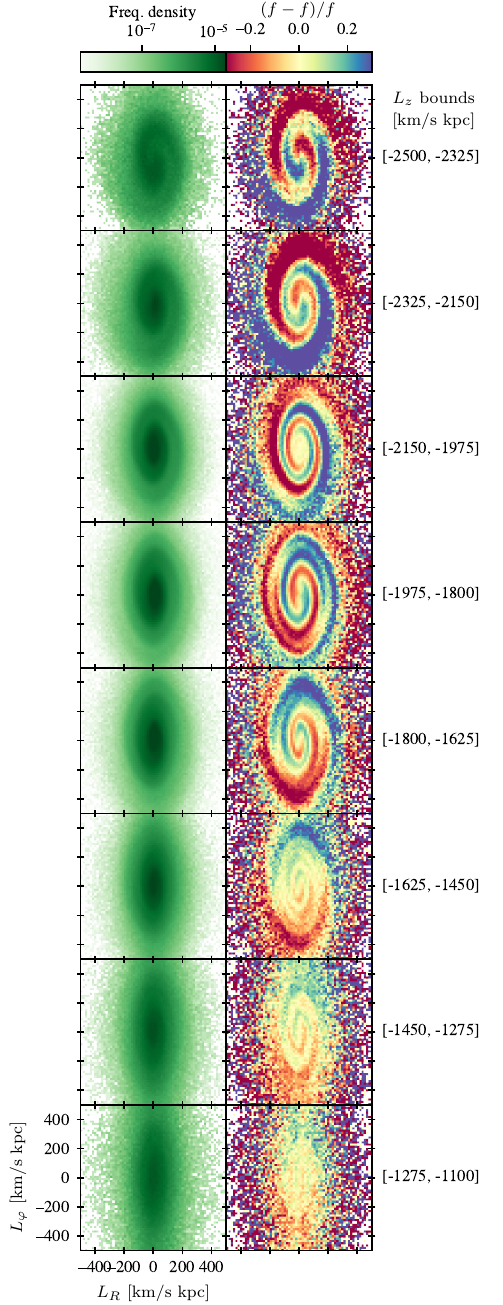}
    \caption{The $L_R-L_\varphi$ spiral split across $L_z$ bins, as labelled. \textit{Left:} raw histograms. \textit{Right:} fractional residuals against symmetrised distributions. The phase, winding, and prominence of the spiral change appreciably with $L_z$.}
    \label{fig:Lzdecomp}
    \end{center}
\end{figure}

Figure \ref{fig:observedspiral} exhibits the main result of this Article: the clear spiral structure visible in the distribution of \textit{Gaia} stars in AM space. In the figure we display the cylindrical coordinate system $L_R, L_\varphi$ where the quantities are given (in terms of cylindrical velocities $v_R, v_\varphi, v_z$) by
\begin{equation}
\label{eq:AMCylindrical}
    \begin{aligned}
        L_R &\equiv -zv_\varphi;\\
        L_\varphi &\equiv zv_R - R v_z.
    \end{aligned}
\end{equation}
The left panel shows the raw histogram of stars in our sample. Even before any processing of the images, there is a clear spiral structure visible. 

In the right panel, we pick out this spiral signal using a basic yet effective procedure: we construct a `symmetrised' histogram $\tilde{f}$ by taking the average of the original histogram $f$ and its 180-degree rotation. The right panel then shows the fractional residuals of the original histogram versus this symmetrised histogram, i.e., $f/\tilde{f} - 1$. This procedure is designed to extract any odd-parity signal in the histogram, but is not perfect: there might well be even-parity structure that is washed away, and it is highly sensitive to the position of the zero point (in other words, sensitive to the choice of mid-plane). These caveats notwithstanding, the method appears to successfully extract the spiral structure from the stellar distribution: the residuals clearly show spiral-shaped over/under-densities at around a $\sim 20\%$ level.

We also show, through simple arguments, that the AM spiral shown in Figure \ref{fig:observedspiral} relates to the $z-v_z$ spiral through simplifying assumptions about the local volume. 
To show the mapping of the $z-v_z$ spiral to $L_x-L_y$, starting from the definitions of AM $\Vec{L} = \Vec{r} \times \Vec{v}$, considering only $L_x, L_y$ and solving for $z,v_z$ the following relations between AM and phase space coordinates are found
\begin{equation}
\label{eq:z-vz}
    \begin{aligned}
        z &= \frac{-x L_x - y L_y}{\Lz} \\
        v_z &= \frac{-v_x L_x -v_y L_y}{\Lz}
    \end{aligned}
\end{equation}
We can further simplify the relations by assuming that, locally, $v_y \gg v_x$, and $x \gg y$ and $\Lz \approx xv_y$. Substitution of these relations into Equations \ref{eq:z-vz} yields the local mapping between the phase space spiral and the AM spiral
\begin{equation}
\label{eq:z-vz-local}
    \begin{aligned}
        z_{\rm local} &\approx -\frac{L_x}{v_y}\\
        v_{\rm z, local} &\approx -\frac{L_y}{x}
    \end{aligned}
\end{equation}
The resulting relations for $L_x-L_y$ map to $L_R-L_\varphi$ through $L_x \approx -L_R$ and $L_y \approx - L_\varphi$.\footnote{The definition of the coordinate system implies that at the solar location all stars have $x\approx-R$, $y\approx 0$.}

From these considerations, it is clear that {\it locally} a spiral in $L_R-L_\varphi$ will map to a $z-vz$ spiral. However, this mapping is evident only when making an assumption about the motions in the local volume. Furthermore, this approximation can be applied at various locations on the disc and is valid only if coordinate axes are rotated such that the new frame of reference has its $x$-axis aligned with the galactic centre. 

Figure~\ref{fig:Lzdecomp} shows the $L_R-L_\varphi$ spiral after segregating our stellar sample into a series of bins in vertical AM $L_z$. The spiral structure is discernible in all but the last $L_z$ bin, albeit with differing morphologies. Progressing from largest (most negative) to the smallest (least negative) $\Lz$, the phase of the spiral changes and its winding tightens.

Whether the disappearance of the spiral in the final $L_z$ bin is real or instead due to the paucity or imprecision of the data is unclear. Some asymmetry is discernible in the residuals, but this can be made to disappear by changing the assumed position of the galactic mid-plane.

\section{Interpreting the spiral}
\label{sec:theory}

In this section, we try to arrive at a dynamical interpretation of the AM spiral described in the previous section. 
First, in Sec.~\ref{sec:theory:AMpath}, we consider the path of a typical disc star through AM space, and thus show how differential rotation in this space gives rise to the spiral. Then, in Sec.~\ref{sec:theory:model}, we write down a generative model that can give a good first approximation to the observed spiral.

\subsection{A star's path through angular momentum space}

\label{sec:theory:AMpath}
\begin{figure*}
    \begin{center}
    \includegraphics{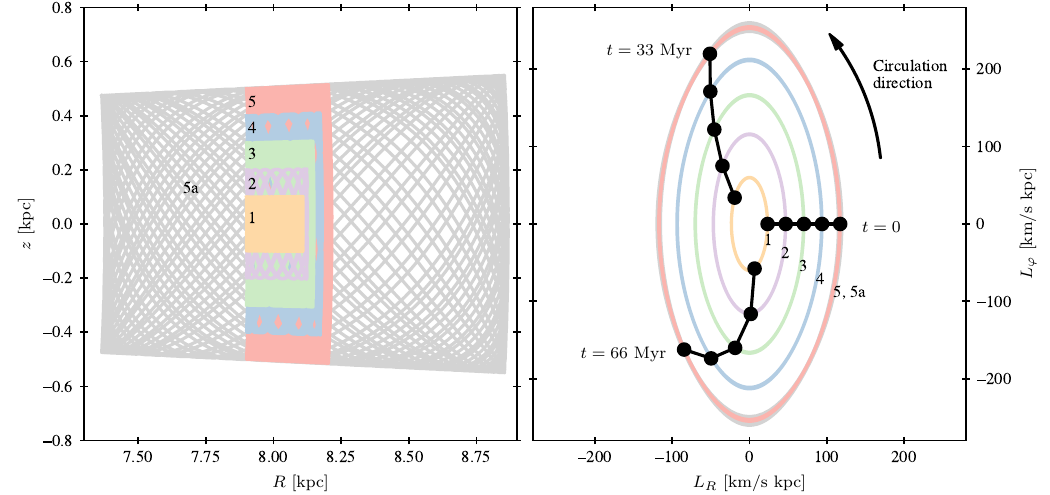}
    \caption{Disc orbits in AM space. \textit{Left:} the trajectories in the meridional plane of 6 stars, labelled `1', `2', `3', `4', `5' (various colours), and `5a' (grey). All stars have the same vertical AM $L_z$ but differ in their vertical energies with the exception of `5a', which has the same vertical energy as `5', but with a much larger radial oscillation. \textit{Right:} the corresponding $L_R-L_\varphi$ trajectories of the same stars (coloured ellipses). Also shown in this panel are isochrones at $t=0$, 33~Myr, 66~Myr (black, spotted lines). The differential circulation of stars in this space gives rise to the observed spiral structure.}
    \label{fig:orbits}
    \end{center}
\end{figure*}

To understand the source of the observed spiral in AM space, we must first understand how typical disc stars on nearly circular orbits move in this space. Taking a single such star with vertical AM $L_z$, we can write down its trajectory in the meridional plane as
\begin{equation}
\label{eq:orbitRz}
    \begin{aligned}
        R &= R_g \left(1 + \epsilon \sin(\Omega_R t)\right) ;\\
        z &= z_0 \sin(\Omega_z t),
    \end{aligned}
\end{equation}
where $R_g$ is the guiding-centre radius of the star, $z_0$ is the amplitude of the star's vertical (i.e., perpendicular to the plane) oscillation, $\epsilon$ is the eccentricity of the in-plane orbit, and $\Omega_z, \Omega_R$ are the respective frequencies of the vertical and radial oscillations. We have assumed that both oscillations have zero initial phase. Including non-zero phases in the sinusoids would lead to no qualitative change in our subsequent results, only phase shifts. These expressions also assume a harmonic and separable potential, but as we shall see, they give a good approximation to the motion in a more realistic potential. 

The guiding-centre radius $R_g$ is a function of $L_z$ alone, given implicitly by $L_z = -v_\mathrm{circ}(R_g)R_g$, where $v_\mathrm{circ}(R_g)$ is the MW circular velocity curve evaluated at $R=R_g$. The negative sign here corresponds to a clockwise (left-handed) rotation around the Galactic centre. Meanwhile, the vertical frequency $\Omega_z$ is a function of the vertical amplitude $z_0$: $\Omega_z \approx 2\pi / P(z)$, with $P(z)$ the vertical period given by
\begin{equation}
    \label{eq:period}
    P = 4\int_0^{z_0} \frac{dz}{\sqrt{2[\Phi(R_g, z_0) - \Phi(R_g, z)]}}.
\end{equation}
In other words, stars with large vertical amplitudes have slower vertical frequencies in a typical Galactic potential.

Derivatives of Equation~(\ref{eq:orbitRz}) along with the fact that $L_z = Rv_\varphi$, lead to expressions for the star's velocities in cylindrical coordinates,
\begin{equation}
\label{eq:orbitvels}
    \begin{aligned}
        v_R &= \epsilon R_g \Omega_R \cos(\Omega_R t);\\
        v_z &= z_0 \Omega_z \cos(\Omega_z t);\\
        v_\varphi &= -\frac{R_g \Omega_\varphi}{1 + \epsilon \sin(\Omega_R t)}.
    \end{aligned}
\end{equation}
Here, $\Omega_\varphi$ is the angular frequency of the MW rotation \textit{at the guiding-centre radius}, i.e.,  $\Omega_\varphi \equiv v_\mathrm{circ}(R_g) / R_g$.

Finally, Eqs.~\ref{eq:orbitRz} and \ref{eq:orbitvels} can be combined to give expressions for the cylindrical angular momenta,
\begin{equation}
    \begin{aligned}
        L_R &\equiv -zv_\varphi\\
        &= \frac{\Omega_\varphi}{\Omega_z} \frac{1}{1 + \epsilon \mathrm{s}_R(t)} L_0 \mathrm{s}_z(t);\\
        L_\varphi &\equiv zv_R - R v_z\\
        &= -L_0 \mathrm{c}_z(t) + \epsilon \left(\frac{\Omega_R}{\Omega_z}\mathrm{s}_z(t)\mathrm{c}_R(t) - \mathrm{c}_z(t)\mathrm{s}_R(t)\right), \\
    \end{aligned}
\end{equation}
where we have abbreviated $\mathrm{s}_X(t) \equiv \sin(\Omega_X t), \mathrm{c}_X(t) \equiv \cos(\Omega_X t)$, and $L_0 \equiv R_g z_0 \Omega_z$. For stars on nearly circular orbits, such that $\epsilon \ll 1$,
\begin{equation}
\label{eq:AMTimeEvolution}
    \begin{aligned}
        L_R &\approx \frac{\Omega_\varphi}{\Omega_z} L_0 \sin(\Omega_zt);\\
        L_\varphi &\approx -L_0 \cos(\Omega_z t). \\
    \end{aligned}
\end{equation}
Thus, in the space of $L_R$ and $L_\varphi$, a star on a nearly circular orbit circulates in an elliptical shape with frequency $\Omega_z$. The semi-major axis of the ellipse is $L_0$ and the major/minor axis ratio is $\Omega_z / \Omega_\varphi$.\footnote{It is nearly always the case that $\Omega_z > \Omega_\varphi$. However, this can be reversed for stars with sufficiently large vertical amplitudes. In such cases, $L_0$ is instead the \textit{semi-minor} axis of the ellipse, and the major/minor axis ratio is $\Omega_\varphi / \Omega_z$. } Empirically, $\Omega_z$ decreases with $z_0$ more slowly than linear, so that $L_0$ increases with $z_0$. In other words, stars with larger vertical oscillations trace out larger ellipses and they do so with lower frequencies. This differential rotation is the process that gives rise to the observed spiral. 

These various processes are depicted in Figure~\ref{fig:orbits}. The left-hand panel of the figure shows the trajectories in the meridional plane of 6 stars, labelled `1', `2', `3', `4', `5', and `5a'. All stars have the same vertical AM $L_z$, corresponding to a guiding centre radius $R_g=8~\mathrm{kpc}$. However, the stars differ in their vertical energies: star `1' has the smallest vertical energy (smallest oscillation amplitude perpendicular to the plane), while star `5' has the largest. Star `5a' has the same vertical energy as star `5', but with a larger amplitude radial oscillation. The orbits were integrated using the {\tt gala} python package \citep{gala} using the {\tt MilkyWayPotential2022} potential. 

The right-hand panel of Fig.~\ref{fig:orbits} shows the corresponding trajectories in $L_R-L_\varphi$. The stars trace out clear ellipses in this space, as predicted by Eq.~\ref{eq:AMTimeEvolution}, with the ellipse radius increasing monotonically with vertical energy. Comparing stars `5' and `5a', the vastly differing radial oscillation amplitudes leave little imprint in this space: the ellipse corresponding to `5a' is somewhat blurred compared to that of `5' at the vertical extremes (i.e., when $L_R=0$ and $|L_\varphi|$ is maximal). Thus, even stars with non-negligible eccentricities give approximately elliptical traces in $L_R-L_\varphi$.

In this calculation, all stars start at their vertical maxima; this corresponds to the black spotted line labelled `$t=0$' in the right-hand panel of Fig.~\ref{fig:orbits}. The lines labelled `$t=33\ \mathrm{Myr}$' and `$t=66\ \mathrm{Myr}$' can be understood as later isochrones: they indicate the positions of the stars at later instants of time, as labelled (note that star `5a' is not included in this spotted line, for visual clarity). As time goes on, the isochrones grow increasingly curved because the stars with larger vertical energies oscillate more slowly. This illustrates the principle outlined above: the differential circulation in $L_R, L_\varphi$ gives rise to the observed spiral.

\subsection{A generative model for the spiral}
\label{sec:theory:model}

\begin{figure*}
    \begin{center}
    \includegraphics[scale=0.85]{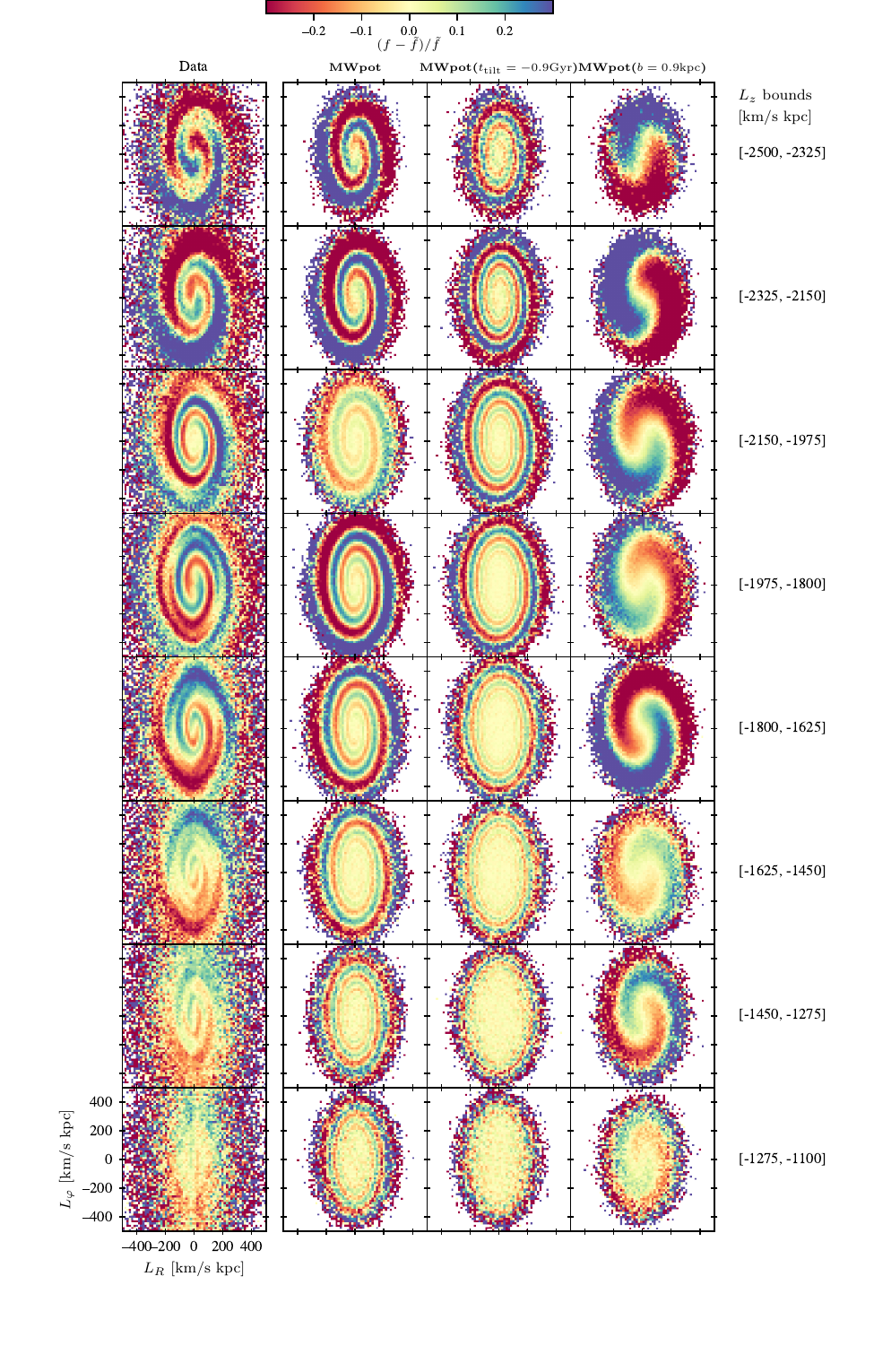}
    \caption{The AM residuals in $L_R-L_\varphi$ binned in $L_z$. \textit{First column:} The {\it Gaia} data.
    \textit{Second column:} A realisation of the MWpot model defined in \ref{sec:theory:model}, sampled at the location of the data. \textit{Third column:} A realisation of the MWpot model with $t_{\rm tilt}=-0.9$ Gyr, the larger time shows a more wound spiral in all bins in comparison to the first model with $t_{\rm tilt}=-0.45$ Gyr. \textit{Fourth column:} A realisation of the MWpot model with the same timing as the first column, but with the scale height of the disc increased to $b= 0.9 \kpc$, this model with a larger scale height shows a much less wound spiral.}
    
    \label{fig:toymodel}
    \end{center}
\end{figure*}

In this section we describe a simple model for the AM spiral. Rather than $L_R, L_\varphi$, modelling is easier in the space $L_R, L'_\varphi$, where $L'_\varphi \equiv (\Omega_\varphi / \Omega_z) L'_\varphi$. In this space, stellar orbits trace out circles rather than ellipses, so prior to the perturbation we can assume a rotationally symmetric equilibrium distribution. We assume a 2D Gaussian with standard deviation $\sigma_L$.

In this simplistic first treatment, we assume the perturbation has acted to rigidly tilt the Galactic disc midplane by an angle $\theta_\mathrm{tilt}$ about a line-of-nodes at azimuth $\varphi_\mathrm{tilt}$. Thus, at azimuth $\varphi$ and radius $R$, the height of the Galactic midplane (with respect to the unperturbed height) is
\begin{equation}
    \label{eq:starheightmodel}
    \bar{z}(R, \varphi) = R \sin(\theta_\mathrm{tilt}) \left[\cos(\varphi_\mathrm{tilt}) \sin(\varphi) -  \sin(\varphi_\mathrm{tilt}) \cos(\varphi)\right]
\end{equation}

Assuming all stars are on circular orbits at $R=R_g$ (with $R_g$ itself a function of $L_z$), the change in $L_R$ for a given star is $\delta L_R = - \bar{z}v_\varphi$ and there is no change in $L_\varphi$. We further assume that this perturbation applies instantaneously at time $t_\mathrm{tilt}$. We take $t=0$ for the present day, so that $t_\mathrm{tilt}$ is negative. 

Our model thus has a parameter set $\theta$ comprising 4 free parameters: $\theta = \{\sigma_L, \theta_\mathrm{tilt}, \varphi_\mathrm{tilt}, t_\mathrm{tilt}\}$. For an individual star with observed $L_z, \varphi$, the likelihood of its observed $L_R^0$, ${L'}_\varphi^0$ (the 0 superscripts here indicating the observed values at the present day) is
\begin{equation}
\begin{aligned}
\label{eq:likelihood}
    \ln l(\theta) &\equiv
    \ln p(L^0_R, {L'}^0_\varphi | L_z, \varphi, \theta)\\ 
    &= - \frac{1}{2 \sigma_L^2} \left((L_R^{t_\mathrm{tilt}} - \mu_L)^2 + {{L'}_\varphi^{t_\mathrm{tilt}}}^2  \right) - \ln(2\pi\sigma_L^2),
\end{aligned}
\end{equation}
where
\begin{equation}
\label{E:AMTimeTransformation}
    \begin{aligned}
        L_R^{0} &= L_R^{t_\mathrm{tilt}} \cos(\Omega_z {t_\mathrm{tilt}}) + {L'}_\varphi^{t_\mathrm{tilt}} \sin(\Omega_z {t_\mathrm{tilt}});\\
        {L'}_\varphi^{0} &= - L_R^{t_\mathrm{tilt}} \sin(\Omega_z {t_\mathrm{tilt}}) + {L'}_\varphi^{t_\mathrm{tilt}} \cos(\Omega_z {t_\mathrm{tilt}}), \\
    \end{aligned}
\end{equation}
and
\begin{equation}
    \mu_L= \bar{z} R_g \Omega_\varphi.
\end{equation}

In other words, the present-day distribution of angular momenta is given by a (differential) rotation (Eq.~\ref{E:AMTimeTransformation}) of the initial distribution, assumed Gaussian. The rotation is differential because $\Omega_z$ is itself a function of $L_R^2 + {L'}_\varphi^2$.

In this work, we do not perform a detailed optimisation of this likelihood function in order to fit the observed data. Instead, we simply endeavour to explore how the model responds to its various input parameters by generating a series of model realisations. In each case, we generate a population of stars with $L_z$ and $\varphi$ values matching those of the Gaia stars in our observed sample, but with $L_R$ and $L_\varphi$ randomly drawn from the likelihood function. In order to generate realisations of the models, we first set the model parameters $\theta$ and choose a MW potential. Given the observed values of $L_z, \varphi$ for each star, we compute $R_g$ and $\Omega_\varphi$. Using these quantities, we can calculate the initial $L_R$ displacement for each star, $\mu_L$. Given $\mu_L$ and $\sigma_L$ (the latter being an input model parameter), we can generate sample values for ${L'}_\varphi^{t_\mathrm{tilt}}$ and ${L}_R^{t_\mathrm{tilt}}$ from the initial Gaussian distribution. Each generated `star' can then be associated with a vertical frequency $\Omega_z$, which for a given potential varies monotonically with $({L'}_\varphi^{t_\mathrm{tilt}})^2 + ({L}_R^{t_\mathrm{tilt}})^2$. This $\Omega_z$ is then used to evolve each star to the present day via Equation \ref{E:AMTimeTransformation}.

Figure~\ref{fig:toymodel} shows three example realisations of the model. We have taken the $L_z$ and $\varphi$ values of the observed stars from the data and for any given star, we compute the values of $\Omega_z,\Omega_\varphi$ using the {\it Gala} \citep{gala} package for a given potential. To illustrate the effect of the parameters of the chosen potential from which the frequencies are computed, we compute the model in two potentials. The first being the {\tt MilkyWayPotential} (MWpot in the Figure), which is a 4-component potential model comprised of a nucleus (Hernquist sphere with $M_{\rm nucl} = 1.71\times10^9 M_{\odot}$, $c=0.07 \kpc$), a bulge (Hernquist sphere with $M_{\rm bulge} = 5\times10^9  M_{\odot}$, $c=1.0 \kpc$), disc (Miyamoto-Nagai with  with $M_{\rm disc} = 6.8\times10^{10}  M_{\odot}$, $a=3.0 \kpc$ and $b=0.28 \kpc$) and a halo (NFW with $M_{\rm halo} = 5.4\times10^{11} M_{\odot}$, $r_s=15.62 \kpc$). The second potential is identical to the first with the exception of the scale height of the disc being increased to $b=0.9$~\kpc. 

For the aforementioned examples, we adopt a model parameter set with $t_\mathrm{tilt}=-0.45$, $\varphi_\mathrm{tilt}=-\pi / 4$, $\theta_\mathrm{tilt}=\tan^{-1}(9/ 800)$ (giving a maximal tilt of $90~\mathrm{pc}$ at $R=8~\mathrm{kpc}$), $\sigma_L=80  \kms\kpc$.
We also show in Figure~\ref{fig:toymodel} a model realisation with the first potential, but with a changed value of $t_{\rm tilt}$ from $-0.45$ Gyr to $-0.9$ Gyr.

The figure shows the distribution of sampled angular momenta, segregated into the same $L_z$ bins as in Fig.~\ref{fig:Lzdecomp}. By eye, the MWpot model (second column in Figure \ref{fig:toymodel} appears to give a good match to the observed spiral, with the spiral morphology approximately correct in most $L_z$ bins. It is worth emphasising that this is simply an approximate, `chi-by-eye' fit, and not the result of a more rigorous optimisation, which we reserve for future work. As such, the parameter choice should not be over-interpreted.

In the second column of Figure ~\ref{fig:toymodel}, we show the AM distribution when changing $t_{\rm tilt}$. A 450 Myr difference in the timing shows a spiral with many more wraps than the when $t_{\rm tilt}=-0.45$ Gyr. In many lower $L_z$ bins, the spiral feature at the centre is not resolvable, and in the smallest $L_z$ bin, the spiral feature disappears almost completely. 

Furthermore, when varying the scale height of the disc in the second potential (third column in Figure ~\ref{fig:toymodel}, the resulting spiral (with the same set of perturbation parameters as the first column) shows a very different morphology. The spiral has many fewer wraps compared with the fiducial potential. In the largest $L_z$ bins ($L_z <-2150 \kms\kpc$) the spiral feature has not yet formed. When the scale height of the disc is large, vertical frequency varies slowly with vertical energy, and so the gradient of the differential rotation is shallow and the spiral is slow to form. This is one example of the effect of the choice of MW potential on the morphology of the spiral.
\section{Discussion}
\label{sec:discussion}

\subsection{The global tilt model}
\label{subsec:modelresult}

    Our simple model presented in Section \ref{sec:theory:model} describes the creation and time evolution of a spiral in AM resulting from a bulk tilting of the disc. While Figure \ref{fig:toymodel} is a simple by-eye fit to the observed data (See Figure \ref{fig:Lzdecomp}), it remarkably produces many of the salient features we observe in the data in most $L_z$ bins. In the second column of Fig. \ref{fig:toymodel} the density along the spiral varies smoothly, and does not show any gaps in the relative density along the spiral, or asymmetric features. In the data however, in the bins with $L_z > -1975\kms \kpc$ , the outer wraps in the region of $L_\varphi< 0$ are much more under dense than the model predicts. Furthermore, the model predicts that the spiral becomes more wound with decreasing (less negative) $L_z$, this is also observed in the data up to $L_z =-1625 \kms \kpc$. Here we stress the effect of the choice of zero point (the height of the galactic mid-plane) on the AM density we recover. Small adjustments to the zero point (of order $<100\pc$) can modify the asymmetry in the AM density, leading to a closer match to the model in some $L_z$ bins, and a worse match in others. This suggests that a single mid-plane height at any given $L_z$ (or guiding radius) is too simplistic to describe the data.
    
    The parameters of the MilkyWayPotential model spiral, which we stress should are only by-eye estimates, place the time of the tilting of the disc at $t_{\rm tilt}=-0.4~{\rm Gyr}$, which is consistent with studies \citep{antoja_dynamically_2018,laporteFootprintsSagittariusDwarf2019,liStellarDistributionFunction2021,widmarkWeighingGalacticDisk2022,frankelVerticalMotionGalactic2023,Frankel.ironsnails.2024} for the time of spiral formation. When matching the model to the data, we find also that using a time of $t_{\rm tilt}~ \simeq 1~{\rm Gyr}$ produces a spiral that is too wound in every $L_z$ bin.

    Not only do the perturbation parameters affect the morphology of the spiral, but also the choice of MW potential parameters. This is illustrated clearly in Figure \ref{fig:toymodel}, where increasing the scale height of the disc, while keeping the likelihood parameters fixed produces a spiral that is much less wound. The particular choice of potential affects the constraints that can be placed on $t_{\rm tilt}$ through our modelling. In this potential, the significantly increased vertical scale height of the disc leads to the slower winding of the spiral, due to the slower change in vertical frequency with vertical energy (see Equation \ref{eq:period}). A less-wound spiral can also be produced by choosing a smaller $t_{\rm tilt}$, demonstrating that the scale height and time since perturbation are likely strongly covariant. 

    The remaining model parameters also affect the spiral produced by the model. For $\theta_{\rm tilt}$, adopting larger values (e.g. $\theta_{\rm tilt}~>~2^{\circ}$) results in an increased relative density in regions that are over-dense, and a decrease in regions that are under-dense. This causes a spiral that appears much `sharper', with few regions of relative density close to zero. The parameter $\sigma_L$ controls the spread in relative density of the space. A larger value leads to a blurring between the over/under-dense regions in AM. Finally, $\varphi_{\rm tilt}$ is tied to the spatial evolution of the spiral and whether, in a given $L_z$ bin a spiral will be observed at the solar location, we discuss the spatial and temporal evolution of the spiral in the following section. 
    
\subsection{The timing, time evolution, and different perturbation scenarios}
\label{subsec:differentmodels}

    In Figure \ref{fig:toymodel}, we show the present-day snapshot ($t=0$) of AM spiral. However, this figure does not fully encapsulate the temporal or spatial evolution of the spiral in AM space. We explore these ideas in this section.

    As a function of time, the spiral winds up due to the differential oscillation of stars in AM space. This winding effect can be seen in Figure \ref{fig:toymodel}: the spiral with the earlier perturbation time (third column) is more tightly wound than the spiral with the later perturbation time (second column). In a given $L_z$ bin, the rate of this winding depends, to first order, on the vertical steepness of the MW potential at the guiding-centre radius corresponding to that $L_z$.

    Under this model with a global tilt perturbation, the spiral also shows spatial evolution, due to the spatial variation in the initial perturbation. Stars that were spatially situated along the line-of-nodes at the time of the perturbation did not experience a significant perturbation, while stars at azimuths perpendicular to the line-of-nodes did. At subsequent times, this spatial variation in perturbation amplitude translates to a spatial variation in spiral amplitude. However, because of the (differential) rotation of the Galactic disc, the locus of points with the largest (smallest) spiral amplitudes changes from a line perpendicular (parallel) to the line-of-nodes into a \textit{spatial} spiral in $x, y$. For an observer at a given position, this manifests as a sinusoidal variation in phase spiral amplitude as a function of $L_z$, because stars falling in different $L_z$ bins have different guiding centre radii, and were thus (on average) at different azimuths at the time of the perturbation. This effect is also seen in our alternative model in which the perturbation is modelled as a more localised velocity kick (Appendix~\ref{appendix:kickmodel}).

    Studies on the tilting of the galactic disc due to previous mergers (e.g. Gaia Sausage/Enceladus) have shown that the degree of tilting depends on a multitude of parameters including the mass of the satellite, the initial inclination w.r.t to the disc at infall, the circularity and whether the merger is prograde and retrograde. For mergers with massive satellites (with $M_{\rm sat} = 10^{11} M_{\odot}$), the tilting rate of the disc can be as high as $\sim 8 {\rm ^{\circ} / Gyr}$ but is drastically smaller for satellites with mass of $M_{\rm sat} = 10^{10.5} M_{\odot}$, where the tilting rate decreases to $< 2 {\rm ^{\circ} / Gyr}$ \citep{dodgeDynamicsStellarDisc2022}. Our modelling of the spiral suggests that only small $\theta_{\rm tilt}<2^{\circ}$\footnote{For our MWpot model, the tilting is instantaneous with a tilt angle of  $\theta_{\rm tilt}~\simeq~1.2{\rm ^{\circ}}$.} tilts are needed to produce and AM spiral with similar amplitude to the data. Additionally, the timing of the spiral found point to a relatively recent merger, where evidence suggests that the Sagittarius dwarf may be the main perturber that seeded the AM spiral we observe \citep{antoja_dynamically_2018, laporteFootprintsSagittariusDwarf2019, huntResolvingLocalGlobal2021}. Our model parameters agree more with a light satellite passage than a massive merger scenario. Although we note that our timing results are also dependent on the choice of potential, and ideally both the timing and potential would be fit simultaneously.

    \citet{huntMultiplePhaseSpirals2022} detected multiple arms in the $z-v_z$ phase spiral and suggest multiple perturbations form the multiple arms detected. In their Figure 2, the two-armed spiral is detected in the region of $R_g < 7 \kpc$, assuming a flat rotation curve, this corresponds to a $L_z >= - 1500 \kms\kpc$. In our Figure \ref{fig:Lzdecomp} we do not observe multiple arms in AM. However, It is important to note here that our method of background removal, $(f-\Tilde{f})/\Tilde{f}$, would erase two-armed spiral features. Nevertheless, two armed spirals could be generated in AM through the application of multiple perturbations. For example, a two armed-spiral could form as a result of a breathing mode perturbation induced by the bar \citep{huntResolvingLocalGlobal2021,bennettExploringSgrMilkyWaydisk2022,candlishPhaseMixingDue2014}.

\subsection{Limitations of the model}
    \label{subsec:limitations}

    There are limitations to the model we present in this work. Mainly they are (i) the model requires the assumption of the MW potential, choices of the parameters here will affect predictions made in AM. This limitation may also be advantageous, as the MW potential parameters could be fit to the data. (ii) The assumption of a global tilt is likely too simple, as both simulations \citep[e.g.][]{huangSinkingSatellitesTilting1997,dodgeDynamicsStellarDisc2022} and data on the MW warp \citep[e.g.][]{jonssonTangledWarpMilky2024a} find that discs in galaxies respond to infalling satellites through bending modes \citep{earpTiltingRateMilky2017}. A more sophisticated tilt perturbation model would include a spatially varying warp. (iii) The model assumes that the perturbation has no self gravity,\footnote{A common assumption; but see discussions regarding self gravity in e.g.  \citet{Darling.spiral.2019,Banik.spiral.2022}.} and the vertical frequencies are computed before the onset of perturbation to the disc.
    
    Additionally, an implicit assumption of our model is that the vertical energy of stars is conserved along their orbits, which requires a separable and static potential. This assumption simplifies the dynamics as the vertical frequencies before and after perturbation remain the same, and a time-varying potential is not required. This assumption does not generally hold in more realistic galactic discs. For instance, \citet{solwayRadialMigrationGalactic2012} demonstrated that vertical energy is not conserved during disc evolution, even under relatively weak perturbations such as spiral arms. A more robust quantity like the vertical action is conserved in low-perturbation regimes, as found by \citet{vera-ciroCONSERVATIONVERTICALACTION2016}. In contrast, stronger perturbations, such as those induced by the Galactic bar, degrade both vertical energy and action conservation due to significant vertical heating.

    However, since we assume that the spiral perturbation occurred relatively recently (within the last $<1$ Gyr) and we consider a volume of the disc outside the bar region, this assumption remains reasonable.

\section{Conclusion}
\label{sec:conclusion}
    In this paper, we have presented a novel view of the phase spiral in AM space.  We show in Section \ref{sec:mainresult} the AM spiral observed in {\it Gaia} data in the solar neighbourhood in cylindrical coordinates. We show also that the AM spiral relates to directly to the $z-v_z$ spiral when making assumptions about the local volume, and that a spiral in AM can map to $z-v_z$. In Section \ref{sec:theory:AMpath} we describe fully the path of stars in AM, relating the path in cylindrical AM to the vertical and azimuthal orbital frequencies of stars. In Section \ref{sec:theory:model} we present a simple model of the spiral in AM through modelling orbits in this space where the perturbation is due to a bulk tilting of the disc. In Section \ref{subsec:modelresult} we demonstrate the ability of the model to recover the salient features of the observed spiral in the data across multiple $L_z$ bins, and the affect of timing and potential on the predicted spiral. In an Appendix, we present an additional model of a kick-based perturbation in the framework of the orbit-based model presented in Section \ref{sec:theory:model}, to show the effect of perturbation choice. The main conclusions are: 

    \begin{enumerate}
        \item AM is a novel space in which the phase space spiral can be clearly observed. The advantages of this space over $z-v_z$ are that the dimensions of the problem are coupled and that stars trace out ellipses in AM, simplifying the interpretation of observations.
        \item By representing the path of a star through AM, we find that the mechanism of spiral formation is due to the differential rotation of stars in AM space.
        \item Through our modelling approach, we find that the AM spiral predicted by modelling a tilt perturbation can reproduce the salient features of the observed spiral in {\it Gaia} data, and could be used to constrain MW potential parameters.
        \item Our model (c.f. Figure \ref{fig:toymodel}) shows that the timing of the spiral and vertical profile of the disc are strongly covariant parameters.
        \item The method presented in Section \ref{sec:theory:model} can be used to generate many spiral models under, 1. different types of perturbations (e.g. different interaction geometries between satellite and disc) 2. different MW potentials.
    \end{enumerate}

The sample of stars with measurable radial velocities is expected to increase in {\it Gaia} DR4, sharpening our view of the AM spiral, and emphasising the need for further theoretical development. Future development could also consider additional orbital information, such as metallicity, to try and precisely locate a perturbation time \citep[e.g.][]{Frankel.ironsnails.2024}. The model presented in this work is a step towards unifying the different projections of the local disequilibrium in the MW, and is a promising avenue to try and constrain not only the cause of the perturbation, but also the dynamical structure of the MW.

\section*{Data Availability}
The data used in this paper are available from the \textit{Gaia} archive \url{https://gea.esac.esa.int/archive/}. Our data analysis codes are available upon reasonable request. 

\section*{Acknowledgements}

APN is supported by an Early Career Fellowship from the Leverhulme Trust. MSP acknowledges funding from a UKRI Stephen Hawking Fellowship. This work has made use of data from the European Space Agency (ESA) mission
{\it Gaia} (\url{https://www.cosmos.esa.int/gaia}), processed by the {\it Gaia}
Data Processing and Analysis Consortium (DPAC,
\url{https://www.cosmos.esa.int/web/gaia/dpac/consortium}). Funding for the DPAC
has been provided by national institutions, in particular the institutions
participating in the {\it Gaia} Multilateral Agreement.
We acknowledge and thank the developers of the following software that was used in this work: Gala~\citep{gala}, NumPy~\citep{numpy_harris2020array}, SciPy~\citep{2020SciPy-NMeth}, IPython~\citep{IPYTHON_PER-GRA:2007}, Matplotlib~\citep{mpl_Hunter:2007}, Jupyter~\citep{community_jupyter_2021}.


\bibliographystyle{mnras}
\bibliography{references}

\begin{thebibliography}{}
\makeatletter
\relax
\def\mn@urlcharsother{\let\do\@makeother \do\$\do\&\do\#\do\^\do\_\do\%\do\~}
\def\mn@doi{\begingroup\mn@urlcharsother \@ifnextchar [ {\mn@doi@} {\mn@doi@[]}}
\def\mn@doi@[#1]#2{\def\@tempa{#1}\ifx\@tempa\@empty \href {http://dx.doi.org/#2} {doi:#2}\else \href {http://dx.doi.org/#2} {#1}\fi \endgroup}
\def\mn@eprint#1#2{\mn@eprint@#1:#2::\@nil}
\def\mn@eprint@arXiv#1{\href {http://arxiv.org/abs/#1} {{\tt arXiv:#1}}}
\def\mn@eprint@dblp#1{\href {http://dblp.uni-trier.de/rec/bibtex/#1.xml} {dblp:#1}}
\def\mn@eprint@#1:#2:#3:#4\@nil{\def\@tempa {#1}\def\@tempb {#2}\def\@tempc {#3}\ifx \@tempc \@empty \let \@tempc \@tempb \let \@tempb \@tempa \fi \ifx \@tempb \@empty \def\@tempb {arXiv}\fi \@ifundefined {mn@eprint@\@tempb}{\@tempb:\@tempc}{\expandafter \expandafter \csname mn@eprint@\@tempb\endcsname \expandafter{\@tempc}}}

\bibitem[\protect\citeauthoryear{{Alinder}, {McMillan}  \& {Bensby}}{{Alinder} et~al.}{2023}]{Alinder.spiral.2023}
{Alinder} S.,  {McMillan} P.~J.,   {Bensby} T.,  2023, \mn@doi [\aap] {10.1051/0004-6361/202346560}, \href {https://ui.adsabs.harvard.edu/abs/2023A&A...678A..46A} {678, A46}

\bibitem[\protect\citeauthoryear{Antoja et~al.,}{Antoja et~al.}{2018}]{antoja_dynamically_2018}
Antoja T.,  et~al., 2018, \mn@doi [Nature] {10.1038/s41586-018-0510-7}, 561, 360

\bibitem[\protect\citeauthoryear{Antoja, Ramos, López-Guitart, Anders, Bernet  \& Laporte}{Antoja et~al.}{2022}]{antoja_tidally_2022}
Antoja T.,  Ramos P.,  López-Guitart F.,  Anders F.,  Bernet M.,   Laporte C. F.~P.,  2022, \mn@doi [A\&A] {10.1051/0004-6361/202244064}, 668, A61

\bibitem[\protect\citeauthoryear{{Antoja}, {Ramos}, {Garc{\'\i}a-Conde}, {Bernet}, {Laporte}  \& {Katz}}{{Antoja} et~al.}{2023}]{Antoja.spiral.2023}
{Antoja} T.,  {Ramos} P.,  {Garc{\'\i}a-Conde} B.,  {Bernet} M.,  {Laporte} C.~F.~P.,   {Katz} D.,  2023, \mn@doi [\aap] {10.1051/0004-6361/202245518}, \href {https://ui.adsabs.harvard.edu/abs/2023A&A...673A.115A} {673, A115}

\bibitem[\protect\citeauthoryear{{Banik}, {Weinberg}  \& {van den Bosch}}{{Banik} et~al.}{2022}]{Banik.spiral.2022}
{Banik} U.,  {Weinberg} M.~D.,   {van den Bosch} F.~C.,  2022, \mn@doi [\apj] {10.3847/1538-4357/ac7ff9}, \href {https://ui.adsabs.harvard.edu/abs/2022ApJ...935..135B} {935, 135}

\bibitem[\protect\citeauthoryear{Bennett \& Bovy}{Bennett \& Bovy}{2018}]{bennett_vertical_2018}
Bennett M.,  Bovy J.,  2018, \mn@doi [MNRAS] {10.1093/mnras/sty2813}, 482, 1417

\bibitem[\protect\citeauthoryear{Bennett, Bovy  \& Hunt}{Bennett et~al.}{2022}]{bennettExploringSgrMilkyWaydisk2022}
Bennett M.,  Bovy J.,   Hunt J. A.~S.,  2022, \mn@doi [ApJ] {10.3847/1538-4357/ac5021}, 927, 131

\bibitem[\protect\citeauthoryear{Binney \& Sch{\"o}nrich}{Binney \& Sch{\"o}nrich}{2018}]{binneyOriginGaiaPhaseplane2018}
Binney J.,  Sch{\"o}nrich R.,  2018, \mn@doi [MNRAS] {10.1093/mnras/sty2378}, 481, 1501

\bibitem[\protect\citeauthoryear{Candlish, Smith, Fellhauer, Gibson, Kroupa  \& Assmann}{Candlish et~al.}{2014}]{candlishPhaseMixingDue2014}
Candlish G.~N.,  Smith R.,  Fellhauer M.,  Gibson B.~K.,  Kroupa P.,   Assmann P.,  2014, \mn@doi [MNRAS] {10.1093/mnras/stt2166}, 437, 3702

\bibitem[\protect\citeauthoryear{Cao, Li, Sch{\"o}nrich  \& Antoja}{Cao et~al.}{2024}]{caoRadialWaveGalactic2024b}
Cao C.,  Li Z.-Y.,  Sch{\"o}nrich R.,   Antoja T.,  2024, \mn@doi [APJ] {10.3847/1538-4357/ad7b0e}, 975, 292

\bibitem[\protect\citeauthoryear{Community}{Community}{2021}]{community_jupyter_2021}
Community E.~B.,  2021, Jupyter {Book}, \mn@doi{10.5281/zenodo.4539666}, \url {https://doi.org/10.5281/zenodo.4539666}

\bibitem[\protect\citeauthoryear{{Darling} \& {Widrow}}{{Darling} \& {Widrow}}{2019}]{Darling.spiral.2019}
{Darling} K.,  {Widrow} L.~M.,  2019, \mn@doi [\mnras] {10.1093/mnras/sty3508}, \href {https://ui.adsabs.harvard.edu/abs/2019MNRAS.484.1050D} {484, 1050}

\bibitem[\protect\citeauthoryear{{Darragh-Ford}, Hunt, {Price-Whelan}  \& Johnston}{{Darragh-Ford} et~al.}{2023}]{darragh-fordESCARGOTMappingVertical2023}
{Darragh-Ford} E.,  Hunt J. A.~S.,  {Price-Whelan} A.~M.,   Johnston K.~V.,  2023, \mn@doi [ApJ] {10.3847/1538-4357/acf1fc}, 955, 74

\bibitem[\protect\citeauthoryear{Dodge, Slone, Lisanti  \& Cohen}{Dodge et~al.}{2022}]{dodgeDynamicsStellarDisc2022}
Dodge B.~C.,  Slone O.,  Lisanti M.,   Cohen T.,  2022, \mn@doi [MNRAS] {10.1093/mnras/stac3249}, 518, 2870

\bibitem[\protect\citeauthoryear{Earp, Debattista, Macci{\`o}  \& Cole}{Earp et~al.}{2017}]{earpTiltingRateMilky2017}
Earp S. W.~F.,  Debattista V.~P.,  Macci{\`o} A.~V.,   Cole D.~R.,  2017, \mn@doi [MNRAS] {10.1093/mnras/stx1143}, 469, 4095

\bibitem[\protect\citeauthoryear{Eilers, Hogg, Rix  \& Ness}{Eilers et~al.}{2019}]{eilers_circular_2019}
Eilers A.-C.,  Hogg D.~W.,  Rix H.-W.,   Ness M.~K.,  2019, \mn@doi [ApJ] {10.3847/1538-4357/aaf648}, 871, 120

\bibitem[\protect\citeauthoryear{Frankel, Bovy, Tremaine  \& Hogg}{Frankel et~al.}{2023}]{frankelVerticalMotionGalactic2023}
Frankel N.,  Bovy J.,  Tremaine S.,   Hogg D.~W.,  2023, \mn@doi [MNRAS] {10.1093/mnras/stad908}, 521, 5917

\bibitem[\protect\citeauthoryear{{Frankel}, {Hogg}, {Tremaine}, {Price-Whelan}  \& {Shen}}{{Frankel} et~al.}{2024}]{Frankel.ironsnails.2024}
{Frankel} N.,  {Hogg} D.~W.,  {Tremaine} S.,  {Price-Whelan} A.,   {Shen} J.,  2024, \mn@doi [arXiv e-prints] {10.48550/arXiv.2407.07149}, \href {https://ui.adsabs.harvard.edu/abs/2024arXiv240707149F} {p. arXiv:2407.07149}

\bibitem[\protect\citeauthoryear{{GRAVITY Collaboration} et~al.,}{{GRAVITY Collaboration} et~al.}{2021}]{gravitycollaborationImprovedGRAVITYAstrometric2021}
{GRAVITY Collaboration} et~al., 2021, \mn@doi [A\&A] {10.1051/0004-6361/202040208}, 647, A59

\bibitem[\protect\citeauthoryear{{Gaia Collaboration} et~al.,}{{Gaia Collaboration} et~al.}{2016}]{gaia_collaboration_gaia_2016}
{Gaia Collaboration} et~al., 2016, \mn@doi [\aap] {10.1051/0004-6361/201629272}, \href {https://ui.adsabs.harvard.edu/abs/2016A&A...595A...1G} {595, A1}

\bibitem[\protect\citeauthoryear{{Gaia Collaboration} et~al.,}{{Gaia Collaboration} et~al.}{2023}]{gaia_collaboration_DR3_2023}
{Gaia Collaboration} et~al., 2023, \mn@doi [\aap] {10.1051/0004-6361/202243940}, \href {https://ui.adsabs.harvard.edu/abs/2023A&A...674A...1G} {674, A1}

\bibitem[\protect\citeauthoryear{{Grand}, {Pakmor}, {Fragkoudi}, {G{\'o}mez}, {Trick}, {Simpson}, {van de Voort}  \& {Bieri}}{{Grand} et~al.}{2023}]{Grand.snail.2023}
{Grand} R. J.~J.,  {Pakmor} R.,  {Fragkoudi} F.,  {G{\'o}mez} F.~A.,  {Trick} W.,  {Simpson} C.~M.,  {van de Voort} F.,   {Bieri} R.,  2023, \mn@doi [\mnras] {10.1093/mnras/stad1969}, \href {https://ui.adsabs.harvard.edu/abs/2023MNRAS.524..801G} {524, 801}

\bibitem[\protect\citeauthoryear{Guo, Li, Shen, Mao  \& Liu}{Guo et~al.}{2024}]{guoMeasuringMilkyWay2024}
Guo R.,  Li Z.-Y.,  Shen J.,  Mao S.,   Liu C.,  2024, \mn@doi [ApJ] {10.3847/1538-4357/ad037b}, 960, 133

\bibitem[\protect\citeauthoryear{Harris et~al.,}{Harris et~al.}{2020}]{numpy_harris2020array}
Harris C.~R.,  et~al., 2020, \mn@doi [Nature] {10.1038/s41586-020-2649-2}, 585, 357

\bibitem[\protect\citeauthoryear{Huang \& Carlberg}{Huang \& Carlberg}{1997}]{huangSinkingSatellitesTilting1997}
Huang S.,  Carlberg R.~G.,  1997, \mn@doi [ApJ] {10.1086/303977}, 480, 503

\bibitem[\protect\citeauthoryear{Hunt, Stelea, Johnston, Gandhi, Laporte  \& B{\'e}dorf}{Hunt et~al.}{2021}]{huntResolvingLocalGlobal2021}
Hunt J. A.~S.,  Stelea I.~A.,  Johnston K.~V.,  Gandhi S.~S.,  Laporte C. F.~P.,   B{\'e}dorf J.,  2021, \mn@doi [MNRAS] {10.1093/mnras/stab2580}, 508, 1459

\bibitem[\protect\citeauthoryear{Hunt, {Price-Whelan}, Johnston  \& {Darragh-Ford}}{Hunt et~al.}{2022}]{huntMultiplePhaseSpirals2022}
Hunt J. A.~S.,  {Price-Whelan} A.~M.,  Johnston K.~V.,   {Darragh-Ford} E.,  2022, \mn@doi [MNRAS: Letters] {10.1093/mnrasl/slac082}, 516, L7

\bibitem[\protect\citeauthoryear{Hunt, Price-Whelan, Johnston, McClure, Filion, Cassese  \& Horta}{Hunt et~al.}{2024}]{hunt_radial_2024}
Hunt J. A.~S.,  Price-Whelan A.~M.,  Johnston K.~V.,  McClure R.~L.,  Filion C.,  Cassese B.,   Horta D.,  2024, \mn@doi [MNRAS] {10.1093/mnras/stad3918}, 527, 11393

\bibitem[\protect\citeauthoryear{Hunter}{Hunter}{2007}]{mpl_Hunter:2007}
Hunter J.~D.,  2007, \mn@doi [Computing in Science \& Engineering] {10.1109/MCSE.2007.55}, 9, 90

\bibitem[\protect\citeauthoryear{J{\'o}nsson \& McMillan}{J{\'o}nsson \& McMillan}{2024}]{jonssonTangledWarpMilky2024a}
J{\'o}nsson V.~H.,  McMillan P.~J.,  2024, \mn@doi [A\&A] {10.1051/0004-6361/202449744}, 688, A38

\bibitem[\protect\citeauthoryear{Khoperskov, Di~Matteo, Gerhard, Katz, Haywood, Combes, Berczik  \& Gomez}{Khoperskov et~al.}{2019}]{khoperskovEchoBarBuckling2019}
Khoperskov S.,  Di~Matteo P.,  Gerhard O.,  Katz D.,  Haywood M.,  Combes F.,  Berczik P.,   Gomez A.,  2019, \mn@doi [A\&A] {10.1051/0004-6361/201834707}, 622, L6

\bibitem[\protect\citeauthoryear{Laporte, Minchev, Johnston  \& G{\'o}mez}{Laporte et~al.}{2019}]{laporteFootprintsSagittariusDwarf2019}
Laporte C. F.~P.,  Minchev I.,  Johnston K.~V.,   G{\'o}mez F.~A.,  2019, \mn@doi [MNRAS] {10.1093/mnras/stz583}, 485, 3134

\bibitem[\protect\citeauthoryear{Li \& Widrow}{Li \& Widrow}{2021}]{liStellarDistributionFunction2021}
Li H.,  Widrow L.~M.,  2021, \mn@doi [MNRAS] {10.1093/mnras/stab574}, 503, 1586

\bibitem[\protect\citeauthoryear{Michtchenko, Barros, {P{\'e}rez-Villegas}  \& L{\'e}pine}{Michtchenko et~al.}{2019}]{michtchenkoMovingGroupsOrigin2019}
Michtchenko T.~A.,  Barros D.~A.,  {P{\'e}rez-Villegas} A.,   L{\'e}pine J. R.~D.,  2019, \mn@doi [ApJ] {10.3847/1538-4357/ab11cd}, 876, 36

\bibitem[\protect\citeauthoryear{P\'erez \& Granger}{P\'erez \& Granger}{2007}]{IPYTHON_PER-GRA:2007}
P\'erez F.,  Granger B.~E.,  2007, \mn@doi [Computing in Science and Engineering] {10.1109/MCSE.2007.53}, 9, 21

\bibitem[\protect\citeauthoryear{Price-Whelan}{Price-Whelan}{2017}]{gala}
Price-Whelan A.~M.,  2017, \mn@doi [The Journal of Open Source Software] {10.21105/joss.00388}, 2

\bibitem[\protect\citeauthoryear{Schönrich, Binney  \& Dehnen}{Schönrich et~al.}{2010}]{schonrich_local_2010}
Schönrich R.,  Binney J.,   Dehnen W.,  2010, \mn@doi [MNRAS] {10.1111/j.1365-2966.2010.16253.x}, 403, 1829

\bibitem[\protect\citeauthoryear{Solway, Sellwood  \& Sch{\"o}nrich}{Solway et~al.}{2012}]{solwayRadialMigrationGalactic2012}
Solway M.,  Sellwood J.~A.,   Sch{\"o}nrich R.,  2012, \mn@doi [MNRAS] {10.1111/j.1365-2966.2012.20712.x}, 422, 1363

\bibitem[\protect\citeauthoryear{Tremaine, Frankel  \& Bovy}{Tremaine et~al.}{2023}]{tremaineOriginFateGaia2023}
Tremaine S.,  Frankel N.,   Bovy J.,  2023, \mn@doi [MNRAS] {10.1093/mnras/stad577}, 521, 114

\bibitem[\protect\citeauthoryear{{Vera-Ciro} \& D'Onghia}{{Vera-Ciro} \& D'Onghia}{2016}]{vera-ciroCONSERVATIONVERTICALACTION2016}
{Vera-Ciro} C.,  D'Onghia E.,  2016, \mn@doi [ApJ] {10.3847/0004-637X/824/1/39}, 824, 39

\bibitem[\protect\citeauthoryear{Virtanen et~al.,}{Virtanen et~al.}{2020}]{2020SciPy-NMeth}
Virtanen P.,  et~al., 2020, \mn@doi [Nature Methods] {10.1038/s41592-019-0686-2}, \href {https://rdcu.be/b08Wh} {17, 261}

\bibitem[\protect\citeauthoryear{Widmark, Laporte  \& Monari}{Widmark et~al.}{2022a}]{widmarkWeighingGalacticDisk2022}
Widmark A.,  Laporte C. F.~P.,   Monari G.,  2022a, \mn@doi [A\&A] {10.1051/0004-6361/202142819}, 663, A15

\bibitem[\protect\citeauthoryear{Widmark, Widrow  \& Naik}{Widmark et~al.}{2022b}]{widmarkMappingMilkyWay2022}
Widmark A.,  Widrow L.~M.,   Naik A.,  2022b, \mn@doi [A\&A] {10.1051/0004-6361/202244453}, 668, A95

\makeatother
\end{thebibliography}


\appendix
\section{Gaia query of dataset}
\label{appendix:query}
The full RVS catalogue used in the paper was queried from the {\it Gaia} archive using the query below:
\begin{verbatim}
SELECT 
    source_id, 
    ra,dec, 
    pmra, pmra_error,
    pmdec, pmdec_error,
    pmra_pmdec_corr,
    parallax,parallax_error, 
    radial_velocity, radial_velocity_error
FROM 
    gaiadr3.gaia_source
WHERE 
    ruwe < 1.4
    AND radial_velocity IS NOT NULL
    AND parallax_over_error > 5 
\end{verbatim}

\section{Modelling the spiral with a kick}
\label{appendix:kickmodel}
In this Appendix we describe a velocity-kick perturbation model, in contrast to the tilt model presented in Section \ref{sec:theory:model}. We describe the resulting angular momentum distribution in $L_R, L'_\varphi$, resulting from a velocity kick to the disc.

We assume the perturbation has taken the form of a `kick' of magnitude $\delta_v$ in vertical velocity $v_z$, so that for any given star, only $L_\varphi$ is affected (cf. Eq.~\ref{eq:AMCylindrical}). Assuming all stars are on circular orbits at $R=R_g$ (with $R_g$ itself a function of $L_z$), the change in $L_\varphi$ is $\delta L_\varphi = -R_g \delta_v$. We further assume that this perturbation applies instantaneously at time $t_\mathrm{kick}$ to all stars within an azimuthal `wedge' centred on $\varphi_\mathrm{kick}$ with full-width $\Delta\varphi$. As in the main text, we take $t=0$ for the present day, so that $t_\mathrm{kick}$ is negative. 

\begin{figure*}
    \begin{center}
    \includegraphics[scale=0.85]{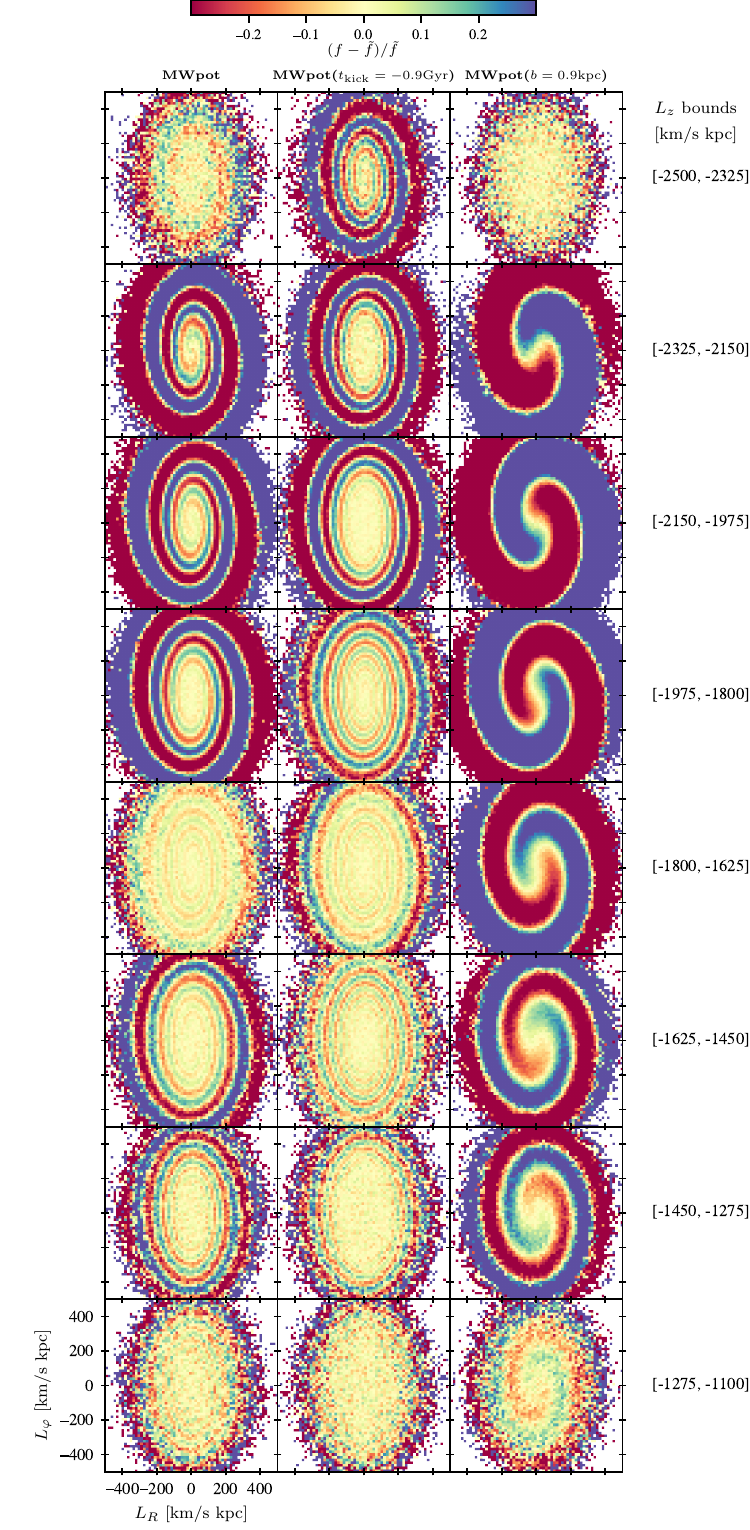}
    \caption{The angular momentum residuals in $L_R-L_\varphi$ binned in $L_z$ for the velocity-kick perturbation model in Appendix~\ref{appendix:kickmodel}. \textit{First column:} A realisation of the MWpot model, sampled at the location of the data. \textit{Second column:} A realisation of the MWpot model with $t_{\rm kick}=-0.9$ Gyr, the larger time shows a more wound spiral in all bins in comparison to the first model with $t_{\rm tilt}=-0.45$ Gyr. \textit{Third column:} A realisation of the MWpot model with the same timing as the first column, but with the scale height of the disc increased to $b= 0.9 \kpc$. In the case of the kick model, the spiral feature appears in less $L_z$ bins.}
    \label{fig:app-kickmodel}

    \end{center}
\end{figure*}

Our model thus has a parameter set $\theta$ comprising 5 free parameters: $\theta = \{\sigma_L, \delta_v, t_\mathrm{kick}, \varphi_\mathrm{kick}, \Delta\varphi\}$. For an individual observed star, the likelihood of its observed $L_R^0$, ${L'}_\varphi^0$ (the 0 superscripts here indicating the observed values at the present day) is
\begin{equation}
\begin{aligned} 
    \ln l(\theta) &\equiv \ln p(L^0_R, {L'}^0_\varphi | L_z, \varphi, \theta)\\
    &= - \frac{1}{2 \sigma_L^2} \left({L_R^{t_\mathrm{kick}}}^2 + ({{L'}_\varphi^{t_\mathrm{kick}}}-\mu_L)^2  \right) - \ln(2\pi\sigma_L^2),
\end{aligned}
\end{equation}
where
\begin{equation}
    \begin{aligned}
        L_R^{0} &= L_R^{t_\mathrm{kick}} \cos(\Omega_z {t_\mathrm{kick}}) + {L'}_\varphi^{t_\mathrm{kick}} \sin(\Omega_z {t_\mathrm{kick}});\\
        {L'}_\varphi^{0} &= - L_R^{t_\mathrm{kick}} \sin(\Omega_z {t_\mathrm{kick}}) + {L'}_\varphi^{t_\mathrm{kick}} \cos(\Omega_z {t_\mathrm{kick}}), \\
    \end{aligned}
\end{equation}
and
\begin{equation}
    \mu_L=
    \begin{cases}
        -\frac{\Omega_\varphi}{\Omega_z}R_g \delta_v, & \text{if $|\varphi - \Omega_\varphi t_\mathrm{kick} - \varphi_\mathrm{kick}| \mod 2\pi < \Delta\varphi / 2$};\\
        0, & \text{otherwise}.
    \end{cases}
\end{equation}
Figure \ref{fig:app-kickmodel} shows the relative angular momentum density in the kick model in the MWpot model used in Section \ref{subsec:modelresult}, as well as the same variations to the time since perturbation and potential as made in the main text. Here the chosen model parameters were $\sigma_L=110 \kms\kpc, \delta_v=14 \kms, \varphi_{\rm kick}=0^{\circ}, \Delta\varphi=280^{\circ}$, for $t_{kick}$ we choose a time of $-0.6$ Gyr (first column in Figure) and $-0.9$ Gyr (second column). 

The choice of a large $\Delta\varphi$ here was made as smaller (i.e. $\Delta\varphi<180^{\circ}$) angles lead to the spiral appearing in half the $L_z$ bin, the appearance (and disappearance) of the spiral in any individual bin depends strongly on this parameter. We arrive at the same conclusions that were made for the global tilt model we presented in the main text. One difference in this kick model is that when observing the stars in a single azimuthal bin, the amplitude of the spiral feature varies much more drastically from one $L_z$ bin to another. This is a consequence of the more localised nature of the initial perturbation. 

\bsp	
\label{lastpage}
\end{document}